\documentclass[sigconf,nonacm,natbib=true,screen=true]{acmart}

\usepackage{color}
\usepackage{bbm}
\usepackage{multirow}

\usepackage[inline]{enumitem}
\usepackage{graphicx}
\usepackage{subcaption}
\usepackage{sistyle}
\usepackage{booktabs}
\usepackage{caption}
\SIthousandsep{,}
\usepackage{makecell}
\usepackage{amsmath}
\usepackage[english]{babel}
\usepackage{hyperref}
\usepackage{array} 
\usepackage{graphicx}
\usepackage{tikz}
\usepackage{wrapfig}
\usepackage{acronym}
\usepackage{graphicx}  
\usepackage{float}  
\usepackage{subcaption}  
\usepackage{fontawesome}
\usepackage[most]{tcolorbox}

\usepackage{algorithm}
\usepackage{algpseudocode}
\newcommand{\heading}[1]{\vspace*{1.1mm}\noindent\textbf{#1.}}

\AtBeginDocument{%
  \providecommand\BibTeX{{%
    \normalfont B\kern-0.5em{\scshape i\kern-0.25em b}\kern-0.8em\TeX}}}

\makeatletter
\g@addto@macro\normalsize{%
  \abovedisplayskip 3pt plus1pt 
  \belowdisplayskip 3pt plus1pt
  \abovedisplayshortskip  0pt plus1pt%
  \belowdisplayshortskip  0pt plus1pt
}
\makeatother

\acrodef{CV}{computer vision}
\acrodef{IR}{information retrieval}
\acrodef{LLM}{large language model}
\acrodef{MDP}{Markov decision process}
\acrodef{NLP}{natural language processing}
\acrodef{NRM}{neural ranking model}
\acrodef{RL}{reinforcement learning}
\acrodef{RL-MARA}{Multi-grAnular Ranking Attack}
\acrodef{MoE}{mixture-of-experts}

\newcommand{\negskip}{\vspace*{-2mm}}

\makeatother


\author{Yuxin Yue}
\authornote{Both authors contributed equally to this research.}
\affiliation{%
  \institution{State Key Laboratory of AI Safety}
  \institution{Institute of Computing Technology, Chinese Academy of Sciences}
  \institution{University of Chinese Academy of Sciences}
  \city{Beijing}
  \country{China}
}
\email{yueyuxin25s@ict.ac.cn}

\author{Yingchen Zhang}
\authornotemark[1]
\affiliation{%
  \institution{State Key Laboratory of AI Safety}
  \institution{Institute of Computing Technology, Chinese Academy of Sciences}
  \institution{University of Chinese Academy of Sciences}
  \city{Beijing}
  \country{China}
}
\email{zhangyingchen23s@ict.ac.cn}

\author{Ruqing Zhang}
\authornote{Corresponding author.}
\affiliation{%
  \institution{State Key Laboratory of AI Safety}
  \institution{Institute of Computing Technology, Chinese Academy of Sciences}
  \institution{University of Chinese Academy of Sciences}
  \city{Beijing}
  \country{China}
}
\email{zhangruqing@ict.ac.cn}

\author{Jiafeng Guo}
\affiliation{%
  \institution{State Key Laboratory of AI Safety}
  \institution{Institute of Computing Technology, Chinese Academy of Sciences}
  \institution{University of Chinese Academy of Sciences}
  \city{Beijing}
  \country{China}
}
\email{guojiafeng@ict.ac.cn}

\author{Maarten de Rijke}
\affiliation{%
  \institution{University of Amsterdam}
  \city{Amsterdam}
  \country{The Netherlands}
}
\email{m.deRijke@uva.nl}

\author{Xueqi Cheng}
\affiliation{%
  \institution{State Key Laboratory of AI Safety}
  \institution{Institute of Computing Technology, Chinese Academy of Sciences}
  \institution{University of Chinese Academy of Sciences}
  \city{Beijing}
  \country{China}
}
\email{cxq@ict.ac.cn}

\renewcommand{\shortauthors}{Yue and Zhang, et al.}

\begin{document}

\title[Faithful Chart Generation for Multimodal Deep Research: Frame--Evidence Co-Adaptation]{Faithful Chart Generation for Multimodal Deep Research: Frame--Evidence Co-Adaptation}

\begin{abstract}
Analytical charts in multimodal deep research encode quantitative claims, requiring every visualized value to be faithfully grounded in supporting evidence. 
Unlike retrieved images that mainly provide contextual information, charts require numerical fidelity: visualized values should not only match retrieved evidence quantitatively but also preserve its original meaning and scope.
However, achieving such fidelity remains challenging because current systems usually construct visualization plans before knowing what quantitative evidence can actually be retrieved from the web. 
As a result, predefined plans may require entities, temporal ranges, or comparison dimensions that the retrieved evidence only partially supports.
Existing approaches mainly address this issue through post-hoc verification after chart plans are fixed, enabling unsupported values to be identified but leaving the underlying visual frames unchanged.
To address this challenge, we propose Frame--Evidence Co-Adaptation (FECA), an evidence-adaptive visual planning framework for multimodal deep research.
Inspired by the bidirectional sensemaking process in Data--Frame Theory, 
FECA models chart generation as an iterative interaction between visual frames and retrieved evidence. 
Each visual frame is adaptive: the frame guides evidence acquisition, while retrieved evidence determines whether the frame should be accepted, revised, or dropped before rendering.
By coupling visualization planning with evidence availability, FECA shifts chart generation from fixed-plan verification to adaptive evidence-grounded visual reasoning. 
Experiments on 100 real-world research topics show that FECA substantially improves numerical fidelity while preserving report quality and chart utility.
\end{abstract}

\maketitle

\section{Introduction}

Deep research systems~\cite{zheng-etal-2025-deepresearcher, chen-etal-2025-mindsearch, ye-etal-2026-deep, chen2025dingtalkdeepresearchunifiedmulti} aim to automate end-to-end knowledge discovery by combining web search \cite{jin2025search, wu2025mmsearch,
yao2026mmdeepresearchsimpleeffectivemultimodal}, reasoning \cite{Search-o1}, and information synthesis \cite{asai2026synthesizing} to produce long-form reports with supporting evidence. 
Early systems mainly focused on textual research reports, improving evidence retrieval, information organization, and citation-aware generation~\cite{shao-etal-2024-assisting,li-etal-2025-webthinker,li-etal-2026-webweaver}. 
Recent multimodal deep research systems extend this capability by automatically generating analytical charts that summarize quantitative information collected from the web~\cite{yang-etal-2026-multimodal-deepresearcher,tian-etal-2026-coggen,ma-etal-2026-tvir}. 
Unlike retrieved images that provide illustrative context, charts function as structured visual evidence: each plotted value implicitly makes a quantitative claim that supports the analytical conclusions of the report. 
Thus, trustworthy multimodal deep research requires not only textual correctness but also reliable grounding of generated visual evidence.

\begin{figure}
    \centering
    \includegraphics[width=1.0\linewidth]{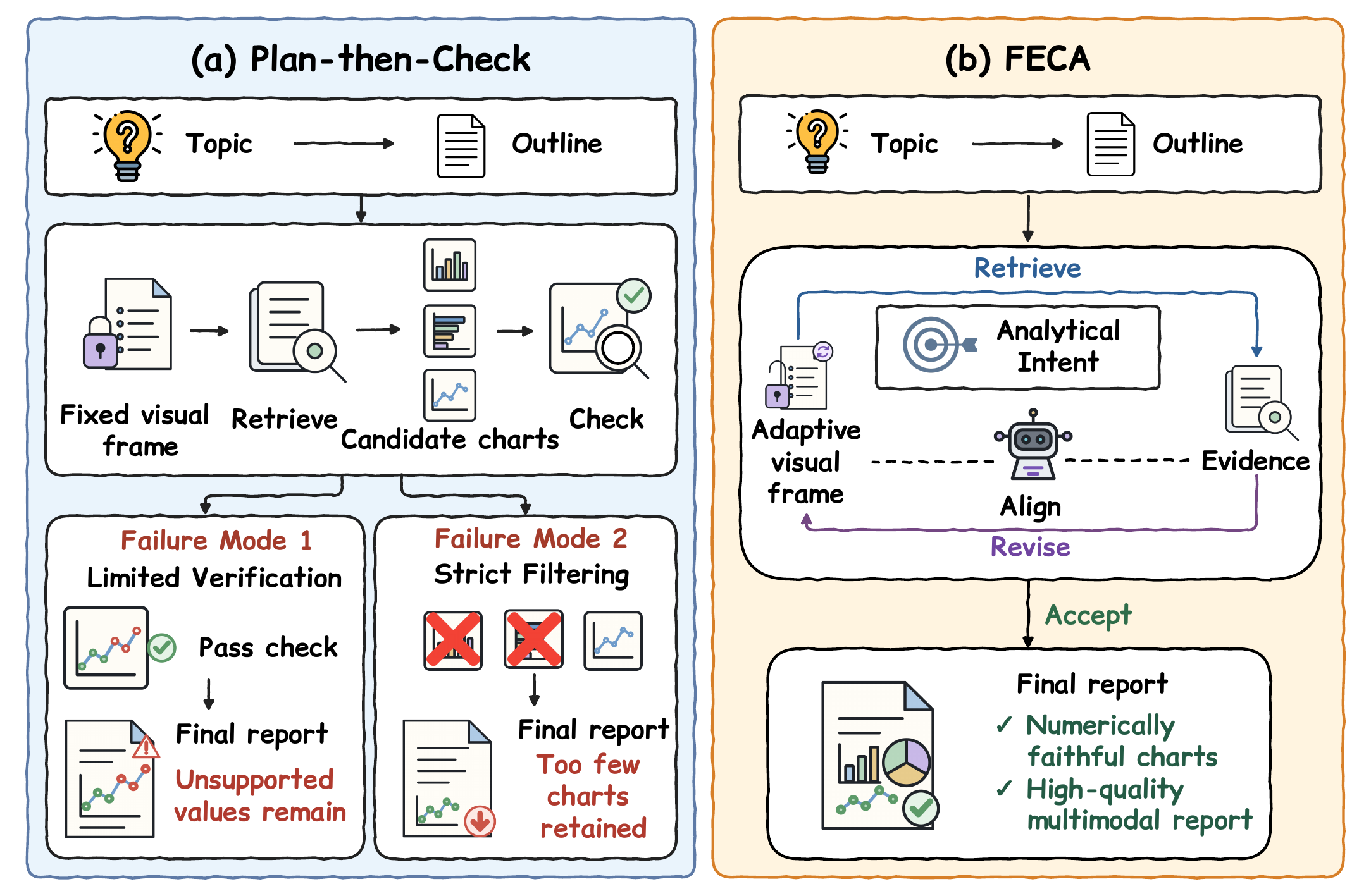}
    \caption{
Unlike fixed visual frames that can only reject unsupported charts, our proposed FECA adapts visual frames according to retrieved evidence to produce evidence-grounded visualizations.
}
    \label{fig:intro}
    \vspace{-2mm}
\end{figure}

\heading{Numerical fidelity: A missing dimension of visual trustworthiness}
Existing evaluations of deep research systems primarily focus on textual report quality and reliability, including information coverage, factuality, and citation correctness~\cite{li-etal-2025-reportbench,du2025deepresearchbenchcomprehensivebenchmark,sharma-etal-2026-researchrubrics}. Recent multimodal extensions further consider visual quality and chart--text consistency~\cite{yang-etal-2026-multimodal-deepresearcher,tian-etal-2026-coggen,ma-etal-2026-tvir}.
However, charts introduce a unique demand of factuality beyond textual claims:
every visualized value corresponds to an implicit numerical assertion that should be traceable to supporting evidence. 

We define this property as \emph{numerical fidelity}:
every value in a generated chart should be grounded in retrieved quantitative evidence while preserving the meaning and scope of its source.
Despite its importance for trustworthy visual evidence~\cite{akhtar-etal-2024-chartcheck,huang-etal-2024-lvlms}, numerical fidelity remains largely unexplored in multimodal deep research.

\heading{The limitation of fixed visualization planning} 
Current multimodal deep research systems typically follow a \emph{plan--then--check} paradigm for chart generation~\cite{yang-etal-2026-multimodal-deepresearcher,ma-etal-2026-tvir,tian-etal-2026-coggen}. 
As illustrated in Figure~\ref{fig:intro}(a), they first instantiate visualization intent as a fixed visual frame, retrieve information required by the predefined frame, and subsequently check whether the retrieved evidence or generated output satisfies that frame.

This paradigm implicitly assumes that a visual frame can be determined before evidence acquisition.
However, web research is inherently uncertain:
retrieved evidence may only partially cover the entities, temporal ranges, or comparison dimensions required by a predefined visual frame.
When such a mismatch occurs, post-hoc verification can revise incorrect values using the available evidence or reject unsupported charts, but cannot adapt the visual frame to the evidence that is actually available. 

\heading{Diagnosing plan--evidence mismatches} 
To understand this limitation, we analyze a representative plan--then--check pipeline in Section~\ref{sec:error-analysis}. 
We find that 80.8\% of its plotted values remain unsupported by the routed evidence after verification.
Standard verification achieves only a limited reduction in the unsupported-value rate.

We further construct a stricter verifier that retains only charts with fully supported values. Although verification improves numerical fidelity, it removes more than three quarters of the generated charts, substantially reducing the visual utility of the resulting reports.
These results reveal a fundamental limitation of verification-based approaches: when a visual frame conflicts with available evidence, stricter checking can only filter invalid outputs rather than discover evidence-supported alternatives. This leads to a key research question: 
\emph{How can multimodal deep research systems generate trustworthy visual evidence when retrieved evidence does not fully satisfy the initial visual frame?}

\heading{Our method: Frame--evidence co-adaptation}
We argue that trustworthy chart generation requires a paradigm shift from verifying fixed visual frames to adapting them according to the evidence actually available. 
As illustrated in Figure~\ref{fig:intro}(b), we propose Frame--Evidence Co-Adaptation (FECA), in which the current visual frame guides evidence acquisition, while retrieved evidence can in turn accept, revise, or drop the frame before chart generation. 
Inspired by the bidirectional sensemaking process in Data--Frame Theory~\cite{klein2007data}, FECA realizes this interaction as an iterative frame--evidence co-adaptation process. 
Data--Frame Theory, a cognitive theory of sensemaking, describes how humans interpret the world under uncertain information through an iterative process: an initial frame guides evidence collection, while newly acquired evidence can confirm, refine, or revise the frame when it conflicts with prior assumptions.

Following this principle, FECA treats each visual frame as adaptive rather than fixed.
Each visual frame specifies a concrete realization of the analytical intent in terms of its target measure, organizing dimensions, and data scope.
The frame guides evidence acquisition, while retrieved evidence determines whether the frame should be accepted, revised, or dropped before chart generation.
Through this bidirectional frame--evidence interaction, FECA enables visual evidence generation that remains aligned with both analytical intent and available web evidence.

Specifically, FECA initializes a candidate visual frame for each
analytical intent identified in the report outline.
During evidence acquisition, it evaluates the compatibility between retrieved evidence and the current frame.
If the evidence sufficiently supports the frame, the chart is directly generated.
If the evidence reveals an alternative visualization that preserves the analytical intent, FECA adapts the frame accordingly.
Otherwise, the unsupported frame is dropped.
By coupling visualization planning with evidence availability, FECA restricts chart generation to visual frames that have been aligned with retrieved quantitative evidence.

\heading{Experimental results}
Experiments on 100 real-world research topics show that FECA substantially improves numerical fidelity over existing systems while maintaining report and chart quality. Ablation and iteration analyses further confirm the importance of evidence-guided frame revision and characterize its effectiveness--cost trade-off. Overall, our findings
show that trustworthy visual evidence generation requires iterative alignment between visual frames and available evidence, rather than verification alone.
\vspace{-2mm}
\section{Our Method}
\label{sec:method}
\begin{figure*}[t]
    \centering
    \includegraphics[width=\textwidth]{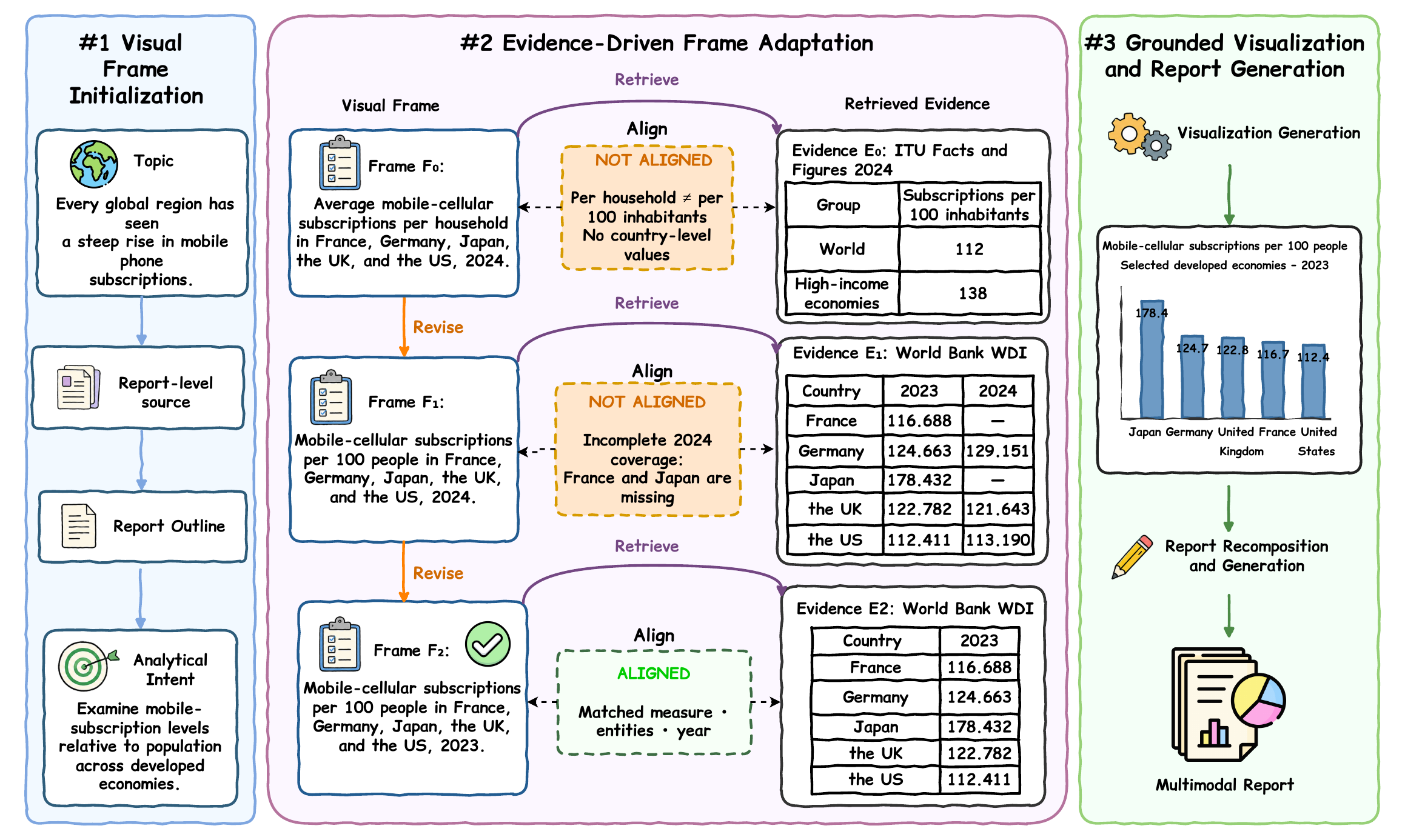}
    \caption{Overview of FECA. Starting from analytical intents, FECA reformulates chart generation as an evidence-adaptive planning process, where visual frames and retrieved evidence iteratively co-evolve: evidence availability determines whether a frame is accepted, revised, or dropped before generating the final evidence-grounded chart.}
    \label{fig:method}
\end{figure*}

In this section, we present Frame--Evidence Co-Adaptation (FECA), a framework for multimodal deep research where visual frames and retrieved web evidence are iteratively co-aligned to produce numerically faithful charts.

\vspace{-2mm}
\subsection{Task formulation and overview}
\label{sec:overview}

Given an open-ended research topic $q$, a multimodal deep research system aims to generate a report $\mathcal{R}$ consisting of textual analysis $\mathcal{T}(\mathcal{R})$ and visual evidence $\mathcal{V}(\mathcal{R})$.
While existing systems primarily optimize the informativeness, coherence, and overall utility of $\mathcal{R}$, we introduce an additional requirement for trustworthy visual evidence: every generated chart should be grounded in quantitatively supported web evidence.

Let $V_i\in\mathcal{V}(\mathcal{R})$ denote a generated chart and $E_i$ its corresponding supporting evidence.
Conceptually, our objective is to maximize the analytical utility of the report while requiring every generated chart to be faithful to its supporting evidence: 
\begin{equation}
\begin{aligned}
\mathcal{R}^{*}
&=
\arg\max_{\mathcal{R}}
Q_{\mathrm{analysis}}(\mathcal{R}) \\
&\quad \text{s.th. }
\operatorname{Faithful}(V_i,E_i)=1,
\ 
\forall V_i\in\mathcal{V}(\mathcal{R}), 
\end{aligned}
\label{eq:our_objective}
\end{equation}
where $Q_{\mathrm{analysis}}(\mathcal{R})$ measures the usefulness of the generated report for answering the research question. 
$\operatorname{Faithful}(V_i,E_i)$ is a binary indicator that equals $1$ when every quantitative value displayed in $V_i$ is traceable to $E_i$ and is consistent with the source evidence.

To realize the above objective, FECA represents each candidate visualization by an analytical intent and an adaptive visual frame.
For visualization $i$, let $I_i$ denote the analytical intent identified from the report outline, which specifies what the visualization is intended to investigate and remains fixed throughout the generation process.
At iteration $t$, $F_i^{(t)}$ denotes the corresponding visual frame, which specifies a concrete way to realize $I_i$ as a chart in terms of its target measure, organizing dimensions, and data scope.
Let $E_i^{(t)}$ denote the quantitative evidence retrieved under the current frame.
During frame--evidence co-adaptation, $F_i^{(t)}$ guides evidence acquisition, while $E_i^{(t)}$ determines whether the current frame can be accepted, should be revised, or must ultimately be dropped.
Thus, the analytical intent remains fixed, whereas its visual frame can adapt to the quantitative evidence available on the web.

As illustrated in Figure~\ref{fig:method}, FECA follows three stages.
It first constructs the report outline and initializes candidate visual frames from analytical intents; it then iteratively retrieves evidence and adapts each frame according to the evidence available; finally, only accepted frame--evidence pairs proceed to grounded chart generation and report recomposition.

\vspace{-3mm}
\subsection{Visual frame initialization}
\label{sec:frame_initialization}

Before retrieving quantitative evidence for individual charts, a research system should first establish the report's analytical objectives.
Otherwise, available data sources may bias the analytical process by prematurely determining what can be analyzed and visualized.
FECA therefore begins with report-level textual research to derive an initial structure, analytical intents, and candidate visual frames that guide subsequent evidence acquisition.

\heading{Report-level source collection}
Given a research topic $q$, we retrieve relevant webpages from the open web and extract their textual content to construct a report-level source set $\mathcal{S}$.
This source set provides broad contextual knowledge for organizing the report and identifying key aspects of the research topic.
We assess whether $\mathcal{S}$ sufficiently covers the major dimensions of $q$ and identify missing aspects that require additional investigation.
For under-covered aspects, we perform an additional retrieval pass and incorporate the newly collected content into $\mathcal{S}$.

\heading{Outline and frame initialization}
Based on $q$ and $\mathcal{S}$, we construct an initial multimodal report outline $\mathcal{O}^{(0)}$.
The outline organizes the report into sections and identifies analytical intents that can potentially benefit from quantitative visual evidence.
For each analytical intent $I_i$, we initialize a visual frame $F_i^{(0)}$, which specifies a preliminary chart design for realizing the intended analysis.
During subsequent frame--evidence co-adaptation, the analytical intent $I_i$ remains fixed, while the corresponding visual frame $F_i^{(0)}$ can be revised or dropped according to the evidence retrieved for it.

At this stage, FECA has established what each visualization is intended to analyze, but the feasibility of realizing these visual frames remains uncertain because the availability of supporting quantitative evidence on the open web is unknown.
The following stage resolves this uncertainty by adapting each visual frame according to the evidence that can be effectively retrieved.

\vspace{-3mm}
\subsection{Evidence-driven frame adaptation}
\label{sec:feca}

FECA treats each visual frame as an adaptive analytical frame rather than a fixed visualization specification.
Given an initial frame $F_i^{(0)}$ and its corresponding analytical intent $I_i$, it performs a bounded \emph{Retrieve--Align--Revise} loop to iteratively adapt the frame according to the available quantitative evidence.
At each iteration, the current frame guides evidence acquisition, and the retrieved evidence determines whether the frame should be accepted, revised, or dropped.
Throughout this process, the analytical intent $I_i$ remains unchanged, while the visual frame  $F_i^{(t)}$ may be revised.

The adaptation process has three outcomes:
(i)~\textsc{Accept}, when the current frame is sufficiently supported by retrieved evidence;
(ii)~\textsc{Revise}, when the evidence reveals a mismatch but an alternative realization of the same intent may exist; or
(iii)~\textsc{Drop}, when no evidence-compatible frame is found within the iteration budget.

\heading{Frame-guided evidence retrieval}
A visual frame specifies the quantitative evidence required to realize an analytical intent, but the existence and availability of such evidence on the open web remain unknown.
At each iteration, \textsc{Retrieve} acquires evidence conditioned on the current frame $F_i^{(t)}$.

Specifically, it searches for candidate data sources that match the target measure, organizing dimensions, and data scope specified by the frame.
Previously discovered sources are ranked and reused when possible; otherwise, new queries are generated to address the remaining evidence gaps.
The retrieved sources are then processed to extract quantitative evidence:
\begin{equation}
E_i^{(t)}
=
\operatorname{Retrieve}
(F_i^{(t)}).
\label{eq:feca_retrieval}
\end{equation}
All quantitative values in $E_i^{(t)}$ are directly copied from their cited sources without modification.

\heading{Frame--evidence alignment}
\label{sec:frame_evidence_alignment} 
The existence of a relevant data source does not necessarily imply that the current frame can be realized.
Retrieved evidence may differ from the frame in terms of target measures, organizing dimensions, or analytical scope.
Therefore, \textsc{Align} evaluates whether $E_i^{(t)}$ sufficiently supports $F_i^{(t)}$:
\begin{equation}
a_i^{(t)}
=
\operatorname{Align}
(F_i^{(t)},E_i^{(t)}),
\quad
a_i^{(t)}\in
\{\textsc{Accept},\textsc{Revise}\}.
\end{equation}
We implement \textsc{Align} as a structured LLM-based evaluation over the frame and retrieved evidence.
The evaluator independently examines whether the target measure, organizing dimensions, and data scope required by $F_i^{(t)}$ are supported by $E_i^{(t)}$.
The frame is accepted only when all required aspects are supported without introducing unsupported quantitative assumptions.
Otherwise, \textsc{Align} returns \textsc{Revise} together with the identified evidence gaps.

Importantly, a revision decision does not indicate that the analytical intent is infeasible.
Instead, the retrieved evidence may support an alternative visual realization of the same intent.

\heading{Evidence-guided frame revision}
\label{sec:frame_revision}
When the retrieved evidence cannot support the current frame, \textsc{Revise} determines whether the mismatch can be resolved without changing the underlying analytical intent. It adapts the current frame according to the retrieved data and previous frame attempts:
\begin{equation}
F_i^{(t+1)}
=
\operatorname{Revise}
\left(
F_i^{(t)},
E_i^{(t)},
\mathcal{H}_i
\mid
I_i
\right),
\label{eq:feca_revision}
\end{equation}
where $\mathcal{H}_i$ records the previous unsuccessful frame attempts for the same analytical intent.

An LLM revises $F_i^{(t)}$ by adjusting its target measure, organizing dimensions, and data scope according to what is supported by $E_i^{(t)}$, while preserving the analytical intent $I_i$. The history $\mathcal{H}_i$ prevents the model from repeatedly proposing frames that have already failed. If a distinct and evidence-compatible frame can be produced, it becomes $F_i^{(t+1)}$ and guides the next retrieval round. If no new frame consistent with $I_i$ can be found, \textsc{Revise} returns $\varnothing$, and the candidate visualization is dropped.

\heading{Termination and output} 
For each candidate visualization, FECA terminates when one of three conditions is met:
(i)~\textsc{Align} returns \textsc{Accept}, producing an accepted frame--evidence pair $(F_i^*,E_i^*)$ (Section~\ref{sec:frame_evidence_alignment});
(ii)~\textsc{Revise} returns $\varnothing$, indicating that no alternative frame consistent with $I_i$ can be constructed (Section~\ref{sec:frame_revision}); or
(iii)~the iteration limit $B$ is reached without an \textsc{Accept} decision. The latter two cases drop the visual frame, while all accepted frame--evidence pairs form the set $\mathcal{C}^*$.

\vspace{-2mm}
\subsection{Grounded visualization and report generation}
\label{sec:constrained_generation}

Once the set of accepted frame--evidence pairs $\mathcal{C}^*$ has been obtained, the task shifts from deciding \emph{what} can be supported by the available evidence to deciding \emph{how} that evidence should be visualized. Detailed visualization specification therefore begins only after a frame--evidence pair has been accepted. Subsequent specification and rendering operate on fixed $F_i^*$ and $E_i^*$ and cannot alter either the frame or its evidence. After rendering, the initial outline is recomposed according to the report-level textual sources, accepted frame--evidence pairs, and successfully rendered visualizations.

\heading{Visualization generation}
For each accepted frame--evidence pair $(F_i^*,E_i^*)\in\mathcal{C}^*$, the visualization planner produces an executable specification $P_i$ that defines the chart form, data bindings, and visual encoding. The renderer then translates $P_i$ into ECharts~\cite{li2018echarts} code:
\begin{equation}
\left(
F_i^*,
E_i^*
\right)
\xrightarrow{\operatorname{Planner}}
P_i
\xrightarrow{\operatorname{Renderer}}
V_i.
\label{eq:visual_generation}
\end{equation}
All data bindings in $P_i$ are derived directly from $E_i^*$. The planner determines how the accepted evidence is visually encoded, but cannot modify its values or introduce quantitative values not contained in $E_i^*$.

A render supervisor then validates the generated chart for specification conformance, successful execution, and visual readability. Rendering failures are returned to the planner or renderer for retries. These retries may correct the specification or implementation, but $F_i^*$ and $E_i^*$ remain unchanged throughout the process. Charts that pass supervision form the final visualization set $\mathcal{V}^*$.

\heading{Report recomposition and generation}
Because frame revision and rendering may change which visualizations are ultimately available, the initial outline $\mathcal{O}^{(0)}$ may no longer accurately reflect the evidence and visualizations produced by FECA. We therefore recompose the outline using the report-level textual sources $\mathcal{S}$, the accepted frame--evidence pairs, and the
successfully rendered visualizations:
\begin{equation}
\mathcal{O}^*
=
\operatorname{Recompose}
\left(
\mathcal{O}^{(0)},
\mathcal{S},
\left\{
\left(
F_i^*,
E_i^*,
V_i
\right)
\right\}_{V_i\in\mathcal{V}^*}
\right).
\label{eq:outline_recomposition}
\end{equation}
\textsc{Recompose} updates the initial narrative claims according to the report-level textual sources, evidence, and rendered
visualizations. In particular, it revises the corresponding section claim to reflect $E_i^*$ and places each visualization alongside the claim it supports.

Finally, the report is generated as
\begin{equation}
\mathcal{R}
=
\operatorname{Write}
\left(
\mathcal{O}^*,
\mathcal{S},
\left\{
\left(
F_i^*,
E_i^*,
V_i
\right)
\right\}_{V_i\in\mathcal{V}^*}
\right).
\label{eq:report_generation}
\end{equation}
The resulting report follows the recomposed narrative and integrates each successfully rendered visualization with its accepted evidence.
\section{Experimental Setup}
\label{sec}
In this section, we describe the dataset, evaluation metrics, baselines, and implementation details used in our experiments.

\negskip
\subsection{Dataset}
\label{sec:dataset}
We evaluate all systems on the 100 research topics in MultimodalReportBench~\cite{yang-etal-2026-multimodal-deepresearcher}, curated from expert-authored multimodal reports published by Pew Research, Our World in Data, and the Open Knowledge Foundation. 

MultimodalReportBench starts from open-ended topics and requires systems to independently research, organize, and visualize information from the web. Topics involve quantitative trends, comparisons, and distributions across heterogeneous sources, making it suitable for evaluating  report utility and numerical fidelity of generated charts. For each topic, we generate a report with each system using its report-generation procedure and evaluate the resulting reports and charts.

\negskip
\subsection{Evaluation metrics}
\label{sec:metrics}
We evaluate the generated reports along three dimensions: numerical fidelity, report and chart quality, and numerical richness.

\heading{Numerical fidelity}
Analogous to atomic-claim evaluation in long-form text~\cite{min-etal-2023-factscore}, we treat each numeric unit displayed in a chart as an independently verifiable quantitative claim and conduct a two-stage audit. The first stage checks each unit against its associated evidence $E_i$ for both value agreement and semantic consistency in measure, unit, entity, and temporal scope. A chart satisfies $\operatorname{Faithful}(V_i,E_i)$ only if all its numeric units pass this audit. Units that cannot be resolved against their associated evidence proceed to a second-stage targeted open-web audit, which searches for external evidence matching the same measure, unit, entity, and temporal scope. This stage distinguishes values that are externally verifiable but unsupported by the evidence originally associated with the chart from values for which no matching evidence can be recovered.

At either stage, after normalizing units and scales, a unit is considered correct only when these semantic attributes match and its relative error from the source value does not exceed 10\%. Each audited unit is assigned one of five mutually exclusive outcomes: Given-Source Correct, Given-Source Error, Open-Search Correct, Open-Search Error, or Not Found. At the chart level, \emph{Provided} measures source availability, while \emph{Valid} measures whether the provided source is accessible and relevant to the chart.

\heading{Report and chart quality}
Following prior evaluations of multimodal research reports~\cite{yang-etal-2026-multimodal-deepresearcher, ma-etal-2026-tvir}, we use an LLM evaluator to assess report-level quality along four dimensions: Coverage and Depth, Coherence and Organization, Visual Integration, and Overall Report Quality. We introduce four chart-level dimensions that assess whether individual charts function effectively as visual evidence:
\begin{enumerate*}[label=(\roman*)]
\item \emph{Chart Contribution} measures whether the chart provides meaningful quantitative evidence for the local analysis.
\item \emph{Chart Self-Documentation} measures whether the chart contains sufficient information to be interpreted independently.
\item \emph{Chart Design} evaluates whether its visual encoding and presentation communicate the underlying data clearly.
\item \emph{Figure--Text Consistency} measures whether the chart is connected to and supports the surrounding discussion.
\end{enumerate*}
All report- and chart-quality metrics are scored on a 1--5 scale.

\heading{Numerical richness}
To determine whether higher fidelity is obtained by displaying less quantitative information, we report Raw Numerical Richness, the average number of numeric units per chart, and Correct Numerical Richness, the average number verified as correct through either audit stage.

\negskip
\subsection{Baselines}
\label{sec:baselines}
We compare FECA with six representative deep research systems from two complementary families.

\heading{Multimodal deep research} 
We compare against three multimodal deep research systems specifically designed to generate text--visual or text--chart interleaved reports.
(i)~\emph{Multimodal DeepResearcher} (MMDR) plans and generates text--chart interleaved reports from open-web research~\cite{yang-etal-2026-multimodal-deepresearcher}.
(ii)~\emph{CogGen} recursively refines report structures and visual plans and audits generated chart coordinates against its collected evidence~\cite{tian-etal-2026-coggen}.
(iii)~\emph{TVIR} uses specialized agents to retrieve images, generate charts, and compose text--visual interleaved reports~\cite{ma-etal-2026-tvir}.

\heading{General-purpose deep research systems}
We further compare against three general-purpose deep research systems that are not specifically designed for multimodal report generation but can conduct open-web research, data analysis, and report synthesis.
(i)~\emph{EvoScientist} is a general-purpose multi-agent research system
whose current open-source implementation coordinates planning, web
research, code execution, data analysis, visualization, and research
writing through specialized agents~\cite{lyu-etal-2026-evoscientist}.
(ii)~\emph{OpenAI Deep Research} (OpenAI-DR) conducts multi-step web research and synthesizes retrieved information into detailed reports with inline citations~\cite{openai2025deepresearch}.
(iii)~\emph{Perplexity Deep Research} (Perplexity-DR) performs extensive web search and synthesizes information from multiple sources into detailed reports with source citations~\cite{perplexity2025sonardeepresearch}.

\negskip
\subsection{Implementation details}
\label{sec:implementation-details}

Across all locally executed systems, including MMDR~\cite{yang-etal-2026-multimodal-deepresearcher}, Cog\-Gen~\cite{tian-etal-2026-coggen}, TVIR~\cite{ma-etal-2026-tvir}, and EvoScientist~\cite{lyu-etal-2026-evoscientist}, we use GPT-4.1-mini for search-query formulation and GPT-4.1 for all other LLM calls in the report-generation pipeline~\cite{openai2025gpt41}. Web search and webpage extraction are performed through the Firecrawl API~\cite{firecrawl2026}. For FECA, MMDR, CogGen, and TVIR, we standardize the initial evidence-collection budget to two retrieval rounds, with three queries issued per round and three results retrieved per query. We preserve each system's original report-generation procedure except for necessary task-specific adaptations. For TVIR, we retain its chart-generation branch but disable its image-retrieval branch, since retrieved images fall outside the scope of our evaluation. For FECA, the maximum frame--evidence co-adaptation budget is set to $B=4$.

OpenAI-DR~\cite{openai2025deepresearch} and Perplexity-DR~\cite{perplexity2025sonardeepresearch} are evaluated through their official APIs and retain their native models, decoding configurations, and retrieval backends. For numerical-fidelity evaluation, we use GPT-5.4~\cite{openai2026gpt54} to extract numerical units from chart images, assess the validity of the provided sources, and verify the extracted units against the corresponding evidence. Firecrawl is used to search, retrieve, and parse cited and open-web sources during verification.

\begin{figure}[t]
\centering
\includegraphics[width=0.85\linewidth]{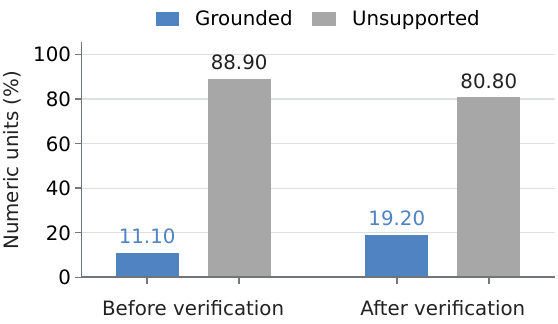}
\caption{Effect of post-hoc verification on the unsupported-value rate in a plan--then--check pipeline.}
\label{fig:verification_mechanism}
\end{figure}

\section{Experimental Results}
\label{sec:experiments}

We organize our experiments around four research questions (RQs).
\textbf{(RQ1)} What limits the plan--then--check paradigm?
\textbf{(RQ2)} Does FECA improve chart-level numerical fidelity without sacrificing report utility?
\textbf{(RQ3)} Which components of FECA contribute to effective frame--evidence co-adaptation?
\textbf{(RQ4)} How does FECA balance adaptation effectiveness and computational cost?

\negskip
\subsection{Limitations of fixed plan--then--check pipelines}
\label{sec:error-analysis}

To answer \textbf{RQ1}, we first investigate whether the numerical fidelity issues in plan--then--check pipelines can be addressed through verification alone.
We evaluate two settings: a practical LLM-based verifier and an idealized oracle filtering mechanism.
Verification can identify unsupported outputs, but cannot resolve mismatches between predefined visual frames and available evidence.

\heading{Practical verification provides limited fidelity improvement}
Following the verification design of CogGen~\cite{tian-etal-2026-coggen}, we implement a representative plan--then--check pipeline that generates charts under fixed visual frames and  verifies plotted values against their corresponding evidence.
The verifier can correct unsupported values when valid supporting evidence is already available; however, it neither modifies the underlying visual frame nor performs additional evidence acquisition.
We randomly sample 20 reports generated by this pipeline and compare charts before and after verification in terms of unsupported-value rate.

As shown in Figure~\ref{fig:verification_mechanism}, verification increases the proportion of supported values from 11.1\% to 19.2\%, but the unsupported-value rate remains high at 80.8\%. These results indicate that post-hoc verification provides only limited improvements in numerical fidelity, leaving most plotted
values unsupported by the routed evidence.

\heading{Oracle verification reveals the limits of fixed frames} 
The limited improvement above raises the question of whether stricter verification could resolve the issue.
To isolate the effect of verification quality, we construct an oracle provenance gate that retains a chart only when every plotted value has a complete provenance chain from the retrieved evidence to the final visualization.

As shown in Table~\ref{tab:oracle-provenance-gate}, oracle filtering reduces the unsupported-value rate to zero by construction, but removes more than three quarters of generated charts and substantially decreases Visual Integration.
This result reveals a fundamental limitation of the plan--then--check paradigm: when visual frames are fixed before evidence acquisition, verification can only reject unsupported charts; it cannot adapt the analytical frame to recover evidence-compatible alternatives.

\begin{table}[t]
\centering
\caption{Oracle provenance filtering eliminates unsupported values but reveals the fidelity--utility trade-off of fixed plan--then--check pipelines.}
\label{tab:oracle-provenance-gate}
\setlength{\tabcolsep}{5pt}
\begin{tabular}{lrrr}
\toprule
Metric & Fixed Plan & Oracle Filter & $\Delta$ \\
\midrule
Unsupported rate (\%) $\downarrow$
& 80.8 & 0.0 & $-80.8$ pp \\
Charts per report $\uparrow$
& 6.00 & 1.45 & $-4.55$ \\
Visual integration $\uparrow$
& 2.82 & 1.73 & $-1.09$ \\
Overall quality $\uparrow$
& 3.09 & 2.91 & $-0.18$ \\
\bottomrule
\end{tabular}
\end{table}

\begin{table*}[t]
\centering
\caption{Source availability and numerical fidelity of generated charts.
Bold and underlined values denote the best and second-best results on fidelity-related metrics, respectively.}
\label{tab:numerical-fidelity}

\setlength{\tabcolsep}{5pt}

\begin{tabular}{l c cc cc ccc}
\toprule
\multirow{2}{*}{System}
& \multirow{2}{*}{\# Charts}
& \multicolumn{2}{c}{Source Availability (\%)}
& \multicolumn{2}{c}{Given-Source Audit (\%)}
& \multicolumn{3}{c}{Open-web Audit (\%)}\\
\cmidrule(lr){3-4}
\cmidrule(lr){5-6}
\cmidrule(lr){7-9}

&
&
Provided $\uparrow$
& Valid $\uparrow$
& Correct $\uparrow$
& Error $\downarrow$
& Correct $\uparrow$
& Error $\downarrow$
& Not Found $\downarrow$\\
\midrule

\textbf{FECA}
& 863
& \underline{99.29}
& \underline{99.29}
& \textbf{82.67}
& \textbf{0.21}
& 3.41
& \textbf{1.37}
& \textbf{12.34} \\

MMDR
& 275
& 82.81
& 67.92
& 8.48
& 1.02
& 1.25
& 12.44
& 76.81 \\

CogGen
& 433
& 26.40
& \textbf{100.00}
& 5.56
& 1.02
& \underline{3.78}
& 8.62
& 81.02 \\

TVIR
& 325
& \textbf{100.00}
& 98.15
& 32.01
& 11.03
& \textbf{5.38}
& 6.81
& 44.77 \\

OpenAI-DR
& 265
& \textbf{100.00}
& 99.15
& \underline{60.73}
& 10.93
& 2.43
& \underline{4.45}
& \underline{21.46} \\

Perplexity-DR
& 663
& \textbf{100.00}
& 98.67
& 55.99
& 4.82
& 2.58
& 5.37
& 31.24 \\

EvoScientist
& 212
& 51.72
& \textbf{100.00}
& 28.43
& \underline{0.97}
& 0.72
& 20.24
& 49.64 \\

\bottomrule
\end{tabular}
\vspace{2mm}
\end{table*}

\negskip
\subsection{Evidence fidelity and report quality}
\label{sec:main-results}

To answer \textbf{RQ2}, we evaluate whether FECA improves evidence-grounded chart generation while preserving the analytical utility of multimodal deep research reports.
We compare FECA with representative baselines in terms of numerical fidelity, report and chart quality, and further conduct a blinded human evaluation to assess whether the observed improvements align with human judgments.

\begin{table*}[t]
\centering
\caption{
Report and chart quality evaluation.
C\&D: Coverage and Depth; C\&O: Coherence and Organization; VI: Visual Integration.
Contrib.: Chart Contribution; Self-Doc.: Chart Self-Documentation; Consist.: Figure--Text Consistency.
Bold and underlined values denote the best and second-best results, respectively.
}
\label{tab:report-quality}

\setlength{\tabcolsep}{4.5pt}

\begin{tabular}{l cccc cccc cc}
\toprule

&
\multicolumn{4}{c}{Report Quality}
&
\multicolumn{4}{c}{Chart Quality}
&
\multicolumn{2}{c}{Numerical Richness}\\

\cmidrule(lr){2-5}
\cmidrule(lr){6-9}
\cmidrule(lr){10-11}

System
& C\&D
& C\&O
& VI
& Overall
& Contrib.
& Self-Doc.
& Design
& Consist.
& Raw
& Correct\\

\midrule

\textbf{FECA}
&3.94
&\textbf{4.02}
&\textbf{3.78}
&\underline{3.89}
&\textbf{3.89}
&\textbf{3.92}
&3.63
&\underline{4.37}
&\textbf{21.05}
&\textbf{18.12}\\

\midrule

MMDR
&\underline{3.97}
&4.00
&3.37
&\textbf{3.90}
&3.58
&3.82
&3.15
&2.77
&13.81
&1.34\\

CogGen
&3.96
&3.89
&2.97
&3.40
&3.04
&3.20
&3.11
&3.45
&\underline{18.84}
&1.76\\

TVIR
&3.94
&3.97
&3.63
&3.87
&3.60
&3.59
&\textbf{3.74}
&4.31
&6.06
&2.26\\

OpenAI-DR
&3.71
&3.99
&\underline{3.66}
&3.82
&\underline{3.71}
&3.07
&3.60
&\textbf{4.42}
&4.22
&2.67\\

Perplexity-DR
&\textbf{4.04}
&\underline{4.01}
&3.45
&3.78
&3.46
&3.04
&3.37
&4.21
&4.02
&2.35\\

EvoScientist
&3.18
&3.96
&2.86
&3.43
&3.31
&\underline{3.84}
&\underline{3.67}
&2.77
&14.31
&\underline{4.17}\\

\bottomrule
\end{tabular}
\vspace{2mm}
\end{table*}

\heading{Numerical fidelity}
We first evaluate whether quantitative claims encoded in generated charts are faithfully grounded in supporting evidence.
Following the audit protocol described in Section~\ref{sec:metrics}, we verify each numeric unit by first checking the source provided with the chart and, when unresolved, conducting targeted open-web searches.
We report source availability and validity at the chart level, as well as the distribution of audit outcomes at the value level.

As shown in Table~\ref{tab:numerical-fidelity}, we observe three findings:
(i) Combining values verified as correct from the provided sources and through open-web search, FECA achieves the highest verified-correct rate of 86.08\% and the lowest \emph{Not Found} rate of 12.34\%. Its verified-correct rate exceeds those of TVIR and OpenAI-DR, the strongest open-source and commercial baselines, by
48.69 and 22.92 percentage points, respectively.
(ii) The advantage of FECA cannot be attributed solely to better source availability.
Several baselines achieve comparable \emph{Provided} and \emph{Valid} scores but substantially lower value-level correctness, indicating that retrieving relevant sources does not guarantee that the quantitative values visualized in charts are actually supported by those sources.
(iii) Additional open-web verification improves FECA's verified-correct proportion by only 3.41 percentage points, suggesting that most correct numeric units are already directly grounded in the evidence retrieved during the FECA process.

\heading{Report and chart quality}
We next investigate whether FECA's fidelity improvement compromises report utility or quantitative richness.
Using the report- and chart-level metrics defined in Section~\ref{sec:metrics}, we evaluate both the overall reports and individual visualizations.
We further distinguish raw numerical richness from correct numerical richness to determine whether improved fidelity is achieved by simply reducing quantitative content.

The results in Table~\ref{tab:report-quality} show that:
(i) At the report level, FECA achieves the best Coherence, Organization, and Visual Integration scores, while remaining within 0.01 and 0.10 points of the best-performing systems on Overall Report Quality and Coverage and Depth, respectively.
This demonstrates that improved numerical fidelity does not come at the cost of reduced analytical coverage or report quality.
(ii) At the chart level, FECA achieves the highest Chart Contribution and Chart Self-Documentation scores, remains within 0.05 points of the best system in Figure--Text Consistency, and maintains competitive Chart Design quality.
These results indicate that FECA-generated charts are not only evidence-grounded but also informative, interpretable, and well integrated with textual analysis.
(iii) FECA provides 11.7\% more numeric units per chart and 4.35$\times$ more verified-correct numeric units than the strongest baseline, showing that fidelity improvements result from better evidence alignment rather than reduced quantitative content.

\begin{table}[t]
\centering
\caption{Human pairwise evaluation over 20 randomly sampled topics.
We report the percentage of topics where FECA wins, ties, or loses against each baseline.}
\label{tab:human-eval}

\setlength{\tabcolsep}{5pt}

\begin{tabular}{lccc}
\toprule
\multirow{2}{*}{Baseline}
& \multicolumn{3}{c}{FECA Win / Tie / Lose (\%)}\\
\cmidrule(lr){2-4}
& Fidelity & Integration & Overall\\
\midrule

MMDR
& \textbf{90/5/5}
& 65/15/20
& 45/20/35\\

TVIR
& \textbf{75/10/15}
& 50/25/25
& 45/25/30\\

OpenAI-DR
& 35/40/25
& 55/15/30
& \textbf{60/10/30}\\

\bottomrule
\end{tabular}
\vspace{-1mm}
\end{table}

\begin{table*}[t]
\centering
\caption{
Ablation study of iterative evidence retrieval and visual frame revision. Bold values indicate best results per column.
}
\label{tab:component-ablation}

\setlength{\tabcolsep}{6pt}

\begin{tabular}{l cc cccccc}
\toprule
& \multicolumn{2}{c}{Components}
& \multicolumn{6}{c}{Results} \\
\cmidrule(lr){2-3}
\cmidrule(lr){4-9}

System
& \shortstack{Iterative\\Retrieval}
& \shortstack{Frame\\Revision}
& \shortstack{\# Charts\\per Report}
& \shortstack{Verified-Correct\\Rate (\%)}
& \shortstack{Visual\\Integration}
& \shortstack{Overall Report\\Quality}
& \shortstack{Chart\\Contribution}
& \shortstack{Figure--Text\\Consistency} \\
\midrule

Single-Pass
& $\times$
& $\times$
& 3.16
& 77.66
& 3.40
& 3.70
& \textbf{3.91}
& 4.18 \\

Fixed Frame
& $\checkmark$
& $\times$
& 4.46
& 70.26
& 3.52
& 3.82
& 3.81
& 4.23 \\

\midrule
FECA
& $\checkmark$
& $\checkmark$
& \textbf{8.63}
& \textbf{86.08}
& \textbf{3.78}
& \textbf{3.89}
& 3.89
& \textbf{4.37} \\

\bottomrule
\end{tabular}
\vspace{1mm}
\end{table*}

\heading{Human evaluation} 
To examine whether the improvements observed in automatic evaluation correspond to human preferences, we conduct a blinded pairwise evaluation on 20 randomly sampled topics from the test set.
We compare FECA against two representative open-source systems (MMDR and TVIR) and the commercial OpenAI-DR.
For each topic, three annotators independently compare two anonymized reports presented in randomized order along three dimensions: Numerical Fidelity, Visual Integration, and Overall Report Quality.
Final preferences are determined by majority voting, and we report the proportion of topics where FECA wins, ties, or loses against each baseline.

As shown in Table~\ref{tab:human-eval}, FECA is consistently preferred over MMDR and TVIR in numerical fidelity and achieves comparable performance with OpenAI-DR.
Across all comparisons, FECA obtains more wins than losses in both visual integration and overall report quality.
These human judgments further confirm that FECA improves numerical grounding while preserving the usefulness and readability of generated research reports.

\negskip
\subsection{Ablation study of frame--evidence co-adaptation}
\label{sec:ablation}

To answer \textbf{RQ3}, we isolate the contributions of iterative evidence retrieval and evidence-guided frame revision, to understand the sources of FECA's improvements. 
We compare the full FECA model with two progressively simplified variants:
(i) \emph{Single-Pass}, which performs only one retrieval round without frame revision; and
(ii) \emph{Fixed Frame}, which enables iterative retrieval but keeps the initial visual frame unchanged.
The full FECA model incorporates both mechanisms to enable iterative frame--evidence co-adaptation.

Table~\ref{tab:component-ablation} shows the impact of each component. (i)~Compared with Single-Pass, Fixed Frame produces 41.1\% more charts and modestly improves Visual Integration and Overall Report Quality. However, its verified-correct rate decreases by 7.40 percentage points, and its Chart Contribution also declines. This suggests that additional retrieval can increase the amount of visualizable evidence, but does not by itself resolve mismatches between a fixed visual frame and the retrieved evidence. (ii)~Adding evidence-guided frame revision to Fixed Frame increases the number of charts per report by 93.5\% and the verified-correct rate by 15.82 percentage points, accompanied by consistent improvements in all report- and chart-quality metrics. Frame revision therefore enables the system to transform newly retrieved evidence into both more abundant and more reliable visual evidence. (iii)~Full FECA achieves the best result on five of the six evaluation metrics and remains within 0.02 points of the best Chart Contribution score. Taken together, these results show that FECA's gains come from combining iterative retrieval with evidence-guided adaptation of the visual frame, rather than from performing additional searches alone.

\vspace{-2mm}
\subsection{Adaptation efficiency and frame resolution dynamics}
\label{sec:adaptation-analysis}

To answer \textbf{RQ4}, we further analyze the efficiency and behavior of FECA during frame--evidence co-adaptation.
Specifically, we study how the iteration budget $B$ affects the quality--cost trade-off and examine how initialized visual frames are resolved throughout the adaptation process.

\heading{Effect of iteration budget on adaptation efficiency}
We examine how the iteration budget affects FECA's adaptation effectiveness and computational cost. We vary the maximum number of iterations $B$ from 1 to 4 while keeping all other settings unchanged. For each budget, we measure the number of charts per report, Visual Integration, Overall Report Quality, and token consumption.

\begin{figure}[t]
    \centering
    \includegraphics[width=\linewidth]{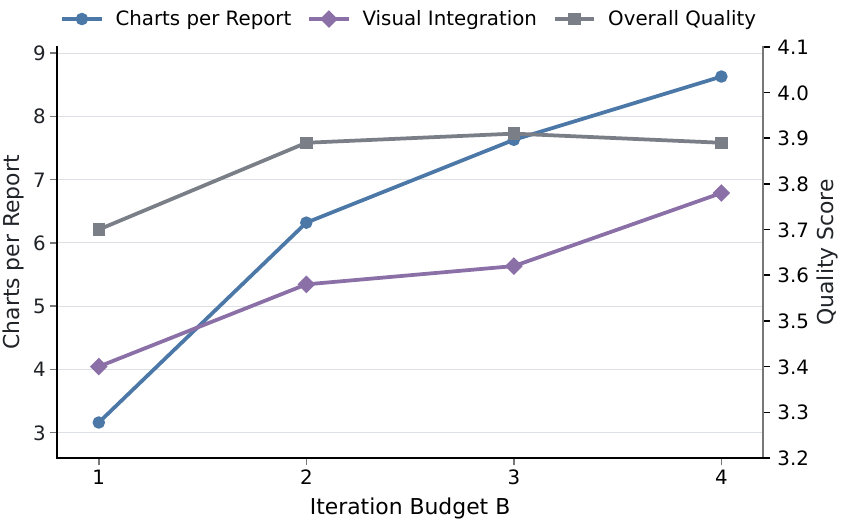}
    \caption{Effect of the iteration budget $B$ on the number of evidence-supported charts and report quality.}
    \label{fig:iteration-budget}
    \vspace*{-1mm}
\end{figure}

Figure~\ref{fig:iteration-budget} shows how the number of evidence-supported charts, Visual Integration, and Overall Report Quality change as $B$ increases. Together with the corresponding token measurements, we observe that (i)~Increasing $B$ consistently produces more evidence-supported charts. The chart count doubles from $B=1$ to $B=2$, while the gains in the third and fourth rounds decrease to 1.31 and 1.00 additional charts per report, respectively. Thus, the second iteration produces the largest increase in the number of evidence-supported charts. (ii)~Visual Integration continues to improve with additional iterations, whereas Overall Report Quality largely saturates after $B=2$. Later iterations therefore primarily improve the availability and integration of visual evidence rather than the overall quality of the report. (iii)~Token consumption increases from 812K at $B=1$ to 1,046K, 1,199K, and 1,279K at $B=2$, $3$, and $4$, respectively. Increasing $B$ from 1 to 4 raises token consumption by 57.5\% but increases the number of charts per report by 173.1\%. These results expose a configurable trade-off: smaller budgets reduce computational cost, whereas larger budgets recover more evidence-supported visualizations. We use $B=4$ in the main experiments to prioritize the generation of evidence-supported charts.

\heading{Frame resolution dynamics during co-adaptation}
We analyze when initialized visual frames become evidence-compatible during the co-adaptation process.
Each frame is categorized according to the iteration at which it is accepted, denoted as \emph{Accept@$b$}, or as \emph{Dropped} if no compatible realization can be obtained within the iteration budget.
\emph{Accept@$1$} corresponds to frames that can be directly supported by retrieved evidence; acceptance in later iterations requires one or more rounds of evidence-guided revision.

\begin{figure}[t]
    \centering
    \includegraphics[width=0.8\linewidth]{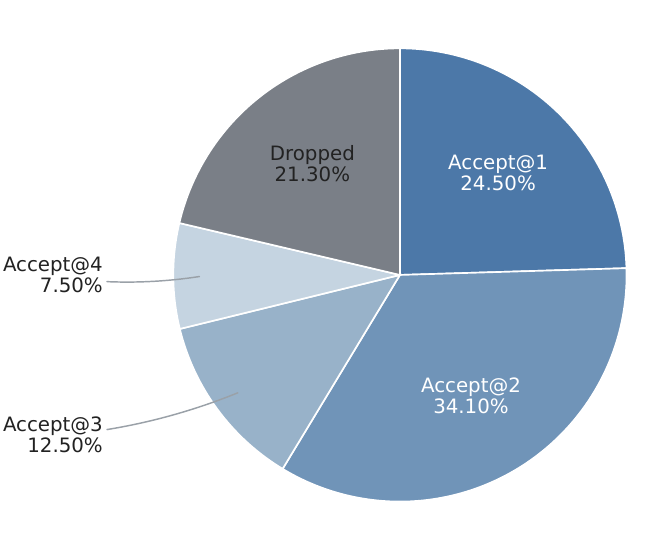}

    \caption{Resolution stages of initialized visual frames during frame--evidence co-adaptation.}
    
    \label{fig:frame-resolution}
\end{figure}

Figure~\ref{fig:frame-resolution} summarizes the resolution process of initialized frames.
We observe that:
(i) only 24.5\% of initialized frames are accepted in the first iteration, while evidence-guided adaptation enables an additional 54.1\% to be accepted in later rounds.
(ii) among all accepted frames, 68.8\% require at least one additional adaptation round, demonstrating that frame revision is a fundamental mechanism rather than an occasional fallback.
(iii) although most frames are accepted within the first two iterations, the third and fourth iterations recover an additional 20.0\% of initialized frames.
Meanwhile, unresolved frames are dropped, preventing unsupported visual claims from entering the final report.
These results explain why increasing the adaptation budget leads to more evidence-grounded visualizations.

\vspace{-2mm}
\section{Related Work}

Reliable chart generation in multimodal deep research lies at the intersection of two research directions: multimodal deep research that aim to improve the trustworthiness of generated reports, and evidence-grounded chart generation methods that study how visualizations can be generated from available data or documents.  

\heading{Multimodal deep research}
Existing multimodal deep research systems apply reliability checks at different stages of chart generation. Multimodal DeepResearcher uses visual feedback to refine generated charts~\cite{yang-etal-2026-multimodal-deepresearcher}; TVIR verifies data authenticity and cross-source consistency before plotting~\cite{ma-etal-2026-tvir}; and CogGen audits generated chart values against its collected evidence~\cite{tian-etal-2026-coggen}. 
Although operating at different stages, these approaches generally assume a fixed visual frame: the frame is determined before knowing whether the open web contains the quantitative evidence required for its realization, and subsequent checks only evaluate whether the retrieved evidence or generated output satisfies the predefined frame.

These methods improve reliability within a fixed visual frame but cannot resolve frame--evidence mismatch when the required evidence is unavailable. FECA addresses this limitation by enabling visual frames to adapt to retrieved evidence before chart generation. 

\heading{Evidence-grounded chart generation}
Outside deep research, reliable chart generation has primarily been studied in settings where the input evidence is known in advance. Existing work can be broadly grouped into two settings: (i) \textit{table-to-chart generation}, where a structured data table and visualization request are provided, ranging from specification generation and constraint-based recommendation~\cite{dibia-demiralp-2019-data2vis, narechania-etal-2021-nl4dv,moritz-etal-2019-draco} to LLM-based methods that improve executable code and visual quality through feedback, debugging, or iterative refinement~\cite{dibia-2023-lida,islam-etal-2024-datanarrative, yang-etal-2024-matplotagent,pesaran-zadeh-etal-2024-text2chart31, koh-etal-2025-c2,wang-etal-2025-dataformulator2}; and (ii) \textit{document-to-chart generation}, where charts are generated from data-bearing documents, either automatically~\cite{zhang2024chartifytext} or according to a predefined chart request~\cite{jain-etal-2025-doc2chart}.  
In both settings, the supporting data or source documents are available before visualization planning and generation, making evidence acquisition outside the scope of the task.

Open-web report generation introduces a different dependency between visualization planning and quantitative evidence acquisition. A system must determine what a chart should communicate without knowing whether the required evidence exists or can be reliably recovered from the web. FECA extends evidence-grounded chart generation to this setting by aligning visual frames with evidence acquired during generation, rather than with evidence provided beforehand. 
\vspace{-2mm}
\section{Conclusion}
In this work, we revisit chart generation in multimodal deep research from the perspective of evidence--visualization alignment. 
We identify a fundamental limitation of existing plan--then--check pipelines: fixed visual frames can only verify or reject generated charts, but cannot adapt when required evidence is unavailable. 
This suggests that trustworthy visualization requires not only validating charts after generation, but also adapting visual frames to the evidence that can actually be obtained. 
To address this challenge, we introduce Frame--Evidence Co-Adaptation (FECA), an evidence-adaptive framework that integrates evidence acquisition with visual frame adaptation. 
FECA preserves high-level analytical intents while allowing their visual realizations to be accepted, revised, or dropped according to retrieved evidence. 
By enabling iterative interaction between visual frames and supporting evidence, FECA reframes chart generation as an evidence-grounded reasoning process rather than fixed-plan verification. 
Experiments on 100 real-world research topics demonstrate that FECA substantially improves numerical fidelity while maintaining the analytical quality and usability of generated reports.

Despite these improvements, several challenges remain: (i) FECA still depends on the availability and reliability of external evidence, and may struggle when relevant quantitative information is sparse, conflicting, or incomplete;   
(ii) FECA relies on LLM-based reasoning for frame--evidence alignment and revision, which can still be affected by hallucinations, uncertainty, or imperfect judgments. As a result, unsupported numerical content may occasionally pass through the co-adaptation process even when the overall procedure operates correctly; and   
(iii) FECA currently keeps the report outline and its analytical intents fixed during frame--evidence co-adaptation, leaving open the broader challenge of jointly adapting higher-level report plans,
visual frames, and evidence during open-ended investigation.  

Future work will explore uncertainty-aware evidence reconciliation, robust numerical verification beyond LLM judgments, and hierarchical co-evolution of research plans, visual frames, and evidence in long-horizon research agents.

\section*{Ethical Considerations}

This work relies on publicly accessible web sources for evidence retrieval and report generation. Third-party sources are used only for research and grounded synthesis, with their original links retained for attribution. We do not intentionally collect private or personally identifiable information. Despite its evidence-alignment mechanisms, FECA cannot guarantee that generated charts are objectively correct: retrieved sources may themselves be outdated, incomplete, or biased, and errors may still arise during retrieval, extraction, frame adaptation, or chart generation. Generated reports should therefore be treated as references requiring human verification rather than as authoritative evidence, particularly in high-stakes domains such as healthcare, finance, and public policy. Our human evaluation involved three graduate-student volunteers who were informed of the study purpose and evaluated anonymized system outputs. The evaluation did not require participants to disclose personal information or intentionally expose them to harmful content.

\bibliographystyle{ACM-Reference-Format}
\balance
\bibliography{references}

\end{document}